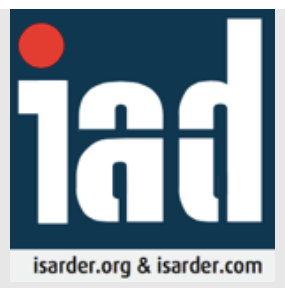

# The Impact of Conspicuous Consumption in Social Media on Purchasing Intentions

**İbrahim Halil EFENDİOĞLU**[a]
[a] Gaziantep University, Department of Informatics, Gaziantep, Turkey, efendioglu@gantep.edu.tr

**ABSTRACT**

**Purpose –** With the rapid increase in the use of social media in the last decade, the conspicuous consumption lifestyle within society has been now transferred to the social media. Along with the changing culture of consumption, the consumer who witnesses such portrayals on social media aspires to and desires the same products and services. Having regard to this situation, this study examines the impact of the conspicuous consumption trend in social media on purchasing intentions. Accordingly, the study aims to discover whether social media is being used as a conspicuous consumption channel and whether these conspicuous portrayals affect purchasing intentions.

**Design/methodology/approach –** A face to face survey was carried out with 409 participants who use social media. The collected data was analyzed with AMOS and SPSS statistical package programs using structural equation modelling.

**Findings –** According to the findings resulting from the analysis, consumer involvement, personal image representation and share satisfaction have been found to have a positive impact on purchasing intention.

**Discussion –** People have started to use social media as a channel for displaying their conspicuous consumption behavior to other people. It can be useful for companies to organize various events on social media and promote their products and services in a customized manner.

## 1. Introduction

Today, 45% of the world's population actively use social media. This number corresponds to 3.5 billion people. They spend 14% of their waking lives on social media (Kemp, 2019). On the other hand, with the increase in the use of social media, a style of conspicuous consumption has been transferred to social media (Bayuk & Öz, 2018: 2846). In fact, sharing posts on social media which reflect conspicuous lifestyles has become a trend (Thoumrungroje, 2014: 8). The different virtual identities created by people on social media legitimize conspicuous consumption. For this purpose, individuals seek public favor by using ostentatious and expensive brands (Sabuncuoğlu, 2015: 370). Conspicuous consumption is the symbolic purchase, retention, use and visible consumption of goods and services equipped with cultural capital, aiming at the communication of a unique image to others. The tendency to conspicuous consumption can be also considered as an innate quality motivating the individual towards different forms of conspicuous consumption, so that an individual can display his/her uniqueness in product selection and use. Thus, conspicuous consumption is one's purchase of products or services to demonstrate one's status or prestige to one's circle of friends or acquaintances (Chaudhuri et al. 2011 :216). Conspicuous consumption was first conceptualized as The Theory of the Leisure Class (Veblen, 1899). In this theory which remains valid today, Veblen suggested that consumption can also be done pretentiously and attributed the relationship between society and economy also to class differentiation (Güleç, 2015: 62). In that sense, people have a tendency to gain status by showing off the products and services they have experienced to others and by using luxury products. In conspicuous consumption, standard solutions are provided by creating artificial needs by the mass media and other constituents of consumption. Besides catering for certain interest groups in society, mass media not only ensures that cultural diversity and social problems are noticed by presenting the lives and experiences of groups and societies, but also it provides information on a social event or phenomenon (Güner Koçak, 2017: 90).

[1]This paper was previously presented as a presentation at the II. Business & Organization Research Conference, September 4-6, 2019, Izmir, Turkey.

The emergence of social media and its evolution into a medium of pretension is an extension of this reality. While consumption in the past was carried out by real people in real life in all its forms and purposes, today it has been moved to a virtual medium, which has begun to become the focus of all displays relating to conspicuous consumption and vanity due to narcissism or motives of public favor or increasing one's power, status and prestige. Social media constitutes this channel of pretension (Bayuk & Öz, 2018 :2846). Today people tend to show by sharing on social media that they are leading a glamorous life (GlobalWebIndex, 2019). Social media, which enables people to communicate with each other, creates a discussion environment and allows participation in social interaction, offers its users the opportunity to conduct more research and evaluate preferences in a short time as a media channel where information is shared and distributed. It is reasonable to assert that trends in consumer behavior are directly related to the use of social media which has a direct impact on consumer behavior (Can, 2017: 209). On the other hand, consumption is not only the purchase of tangible objects, but also a tool for the consumer to display who they want to be. In other words, consumers purchase the goods because of their symbolic meanings (Ünal et al. 2019: 222). Thus, people want to make sure their conspicuous consumption behavior is visible by their circle of friends and their followers. As a consequence of the increase in the number of smart phones and the popularity of the internet, individuals can share their conspicuous consumption on social media. In other words, people gather the luxury goods they have, the interesting places they go, how good they spend their time, the different food they eat under the umbrella of a luxury lifestyle and then tend to show this to people through social media. Driven to legitimize this identity reflected in social media through their consumption, individuals create conspicuous consumption by using glamorous brands which are liked and favored by society and symbolize power, prestige and success (Sabuncuoğlu, 2015: 369-370). In recent years, the majority of luxury brands have begun to more professionally manage their accounts on social media such as on Instagram, Twitter or Facebook, compared to other brands (Statista, 2019).

On the other hand, what is important is the way a consumer evaluates the products or services they intend to purchase. According to the Theory of Consumption Values, a consumer is affected by a number of economic, social and psychological values in their choice of product or service purchase (Sheth et al. 1991). A recent study based on this theory found that social and financial consumption values have a direct impact on purchasing intention, yet emotional and functional consumption values do not have a direct impact on purchasing intention. This finding also indicates that conspicuous consumers prefer the social values such as status, desirability or recognition that they can create by having the product rather than their own feelings and functionality regarding the product. The high price of a product as a financial value makes the consumer feel as if he belongs to a higher social class that purchases expensive products (Ural & Hallumoğlu, 2018: 57). In that sense, purchasing intention represents the consumers' willingness to purchase a certain product in the future and that consumer's plans to purchase the product. An increase in purchasing intention increases the likelihood of the purchase. In order for consumers to develop a purchasing intention regarding a certain brand, they must be able to evaluate all the appropriate brands in the market (Dodds, 1991: 15–17). Since communication between consumers and enterprises has developed thanks to social media, consumers can share their thoughts, requests and their dis/satisfaction concerning a product with both the enterprise and other users through social media. Thus, social media is considered as an element that triggers purchasing intention (Çetin & Kumkale, 2016: 91). On the other hand this research will be useful for luxury brands who want to increase consumer intention to buy using social media.

The main purpose of this research is to determine the impact of "Consumer Involvement", "Personal Image Representation", and "Share Satisfaction", which form the trend towards conspicuous consumption on social media, on purchasing intention. Thus, the impact of the trend towards conspicuous consumption on social media on purchasing intention will be determined and a contribution will be made to the literature on this subject. Therefore, this study which aims at examining the display of conspicuous consumption on social media and its impact on purchasing intention will first explain the important concepts regarding the matter and then investigate previous studies on this subject. Afterwards, the study model, hypotheses, and the supporting arguments will be presented. In the Methodology section, information will be provided regarding data collection and the analysis stage. In the final section, the theoretical and practical contributions of the study will be discussed.

## 2. Literature review

The development of a luxury market in a country goes through five main phases. These are described as: "The Elite", "The Accumulation of Wealth", "The Pretension Period", "Internalization" and "Lifestyle". The elite phase is observed in underdeveloped African countries. There is unbalanced income distribution in this phase and rulers can access a limited number of conspicuous products. The accumulation of wealth phase is observed in BRIC (Brazil, Russia, India and China) countries. There is economic development in this phase. Large numbers of the population spend on items apart from food and clothing items. Elites, on the other hand, prefer luxury. The pretension phase is observed in developing countries. In this phase, enough wealth is accumulated and the middle income group grows the luxury market segment. Luxury is seen as an economic status symbol. The internalization phase is observed in Far Eastern countries. In this phase, luxury is adapted to life and a sophisticated perspective is developed and the supply of luxury goods is diversified. The lifestyle phase is seen in developed countries. In this phase, luxury becomes a part of life and there is a customer mass that knows what they want and can even direct the luxury market (Deloitte, 2018; Gale & Scholz, 1994; Husic & Cicic, 2009) .

### *2.1. Conspicuous Consumption*

Consumer behavior can be defined as a consumers' evaluation before the purchase of certain goods or services, their way of using the goods or services to be purchased as well as their attitudes and behavior following their use of these goods or services. In general, the factors affecting consumer behavior are considered to fall into three categories as follows: demographic factors, socio-cultural factors and psychological factors (Durmaz, 2008: 10–15). It includes all behaviors of individuals during when they use a product or service or prefer or reject an idea or experience to satisfy their needs and desires. Therefore, certain fundamental factors must be discussed to understand why, when and how individuals exhibit their behavior in this process (Solomon, 2010: 27–30). The interest of businesses in the field of marketing where consumption and consumers are at the forefront and their efforts to make themselves noticeable in this field have been gradually increasing. Enterprises continue their race to influence consumers from the manufacturing process to the post-purchase evaluation stage so that they can compete. Consumers, on the other hand, have become much more conscious than ever thanks to advances in technology and have now reached a point where they can follow the market almost as fast as the active manufacturers do and can even direct it. Therefore, much more is needed in this field than what has been done so far in order to understand and impress consumers, who are now almost fully equipped with information (Yazıcı, 2018: 96).

Conspicuous consumption is defined as a type of consumption aimed at ostentation which emerged when consumers, rising to higher levels in the social hierarchy, have increased their demands rather than reducing, despite an increase in product prices and have preferred more expensive goods rather than satisfying their needs with cheaper and more economical goods (Bayuk & Öz, 2018: 2847). Having examined different lifestyles, Veblen considers this new lifestyle as a consumption chain based on the conspicuous consumption theory; and all kinds of products such as clothing, accessories, decorative objects, entertainment and other items can be considered as rings of the chain. He theorizes that people substitute their classical roles in their social lives with their consumption habits when determining their identities. Having inspired postmodernist studies, Veblen's consumption theory has an important place in understanding consumption culture and especially consumer behavior (Adalı Aydın, 2018: 127). Conspicuous consumption is a specific behavior of an individual through which wealth is displayed through luxurious spending on goods and services. Therefore, it differs from general convenience goods due to its prestigious quality and also differs from product satisfaction resulting from user reactions as opposed to real product usage. Consumers purchase certain goods to appear more powerful in the social hierarchy. In mostly capitalist societies, this implies an attempt to achieve a higher social status when considering the consensus nature of conspicuous consumption and the related public meanings(Özdemir & İnanç Sabuncuoğlu, 2018: 10). Today, if consumption is considered as a factor enabling socialization, it is possible to raise awareness in the individual regarding what, where and how to consume during this socialization process. What enables this awareness is mass media and the methods employed by this media such as advertisements, marketing and fashion. In this sense, a socialization process is portrayed when individuals adopt consumption habits, display the products they consume and communicate within their circles through the displayed products within the consumption society. Factors popularizing conspicuous consumption will be discussed within the context of technological advancements, mass media, urbanization, fashion, advertisement and shopping malls (Güner Koçak, 2017: 88). It is defined as a type of consumption aiming at display which emerged when consumers, switching class and climbing to higher levels in the social hierarchy, increased their demand rather than reducing despite an increase in product prices and preferred more expensive goods rather than satisfying their needs with cheaper and more economical goods (Boğa & Başcı, 2016: 465–466).

Among the studies related to conspicuous consumption in the literature, Marcoux et al. (1997) found in their study on the conspicuous consumption behavior of Polish consumers that a tendency to conspicuous consumption is quite common among the youth and Western products are favored over products of other origins. Chen et al. (2005) concluded in their study on the conspicuous consumption habits of Chinese consumers that Chinese immigrants are influenced by the culture they experience in Canada and have a high tendency towards conspicuous consumption. O'Cass and McEwen (2005) assessed the relationship between status consumption and conspicuous consumption. The findings of this study revealed that status consumption and conspicuous consumption are different structures. Hız (2011) determined that many people imitate the upper classes and push their limits by spending more than what they need to achieve status and prestige as well as to display of power. The scale developed by Chaudhuri et al. (2011) to examine the individual differences in conspicuous consumption trends is one of the most important contributions to the literature on conspicuous consumption. Thoumrungroje (2014) investigated the relationship between the frequency of social media use and conspicuous consumption; and his study revealed that the frequency of social media use has a positive, direct and important impact on conspicuous consumption. In addition, it was revealed that people who are more active on social media, in other words those who use social media more intensively, tend to purchase conspicuous consumption products more as a show of prestige; and this can be caused by the users' confidence in electronic word of mouth (EWOM). In their study, Boğa and Başçı (2016) discovered that a tendency to conspicuous consumption varies based on certain demographical characteristics. They showed that conspicuous consumption and monthly personal income level are directly proportional and that individuals who live in societies with high purchasing power are more prone to conspicuous consumption. They found that the higher the individual income attained, the more conspicuous consumption increases. In his study, Güner Koçak (2017) revealed that mass media, advertisements, fashion systems and big shopping malls have an important place in the use of consumed products for display purposes. According to the study, the products consumed provide information about the individuals' economic conditions, lifestyle, social position and status. In their study, Özdemir and İnanç Sabuncuoğlu (2018) identified a meaningful difference in the socio-economic status and conspicuous consumption tendencies of state university students and private university students in Konya. They found that the students from high income groups have a greater tendency towards conspicuous consumption when compared to the students from other income groups. Adalı Aydın (2018) selected 3 television commercials with purposeful sampling and analyzed them using the discourse analysis method. They examined the lifestyles displayed in the text of the advertisement, as well as the consumption-oriented behavior patterns and consumption discourse conveyed to the consumers. It was found that the text of the advertisement presented a lifestyle rather than promoting a product; and instilled common statements such as 'technological advancements', 'global consumption', 'being different/special', 'socializing thanks to the used product', 'change' and 'individualism'. The advertisements examined were observed to reflect the ideology of the consumption culture. Ekşi and Candan (2018) examined the impact of lifestyle on conspicuous consumption. The study concluded that lifestyle directly affects conspicuous consumption.

### *2.2. Conspicuous consumption on social media*

Communication has enabled the lifting of borders in a world shrinking as a result of globalization. Therefore, human values such as speaking, discussing, listening and sharing are able to be rapidly transferred from one side of the world to the other without any loss of time. This development that enables sharing information, emotions and thoughts by linking computer networks has formed the basis of the concept of social media. Deemed as a new lifestyle, perspective and a body of philosophical thought, social media means more than mere transfer. Also considered as a self-realized social transformation project,this concept is a digital habitat where experience, knowledge, thoughts, emotions and judgments are followed, evaluated, interpreted and shared through computer technology (Çetin & Kumkale, 2016: 91). The continuous development of social media and the rapid and global spread of this development has influenced the consumption cultures of consumers as well as many other areas of life. Limited to purchasing products from only physical stores in the past, the consumption culture has mostly, but not entirely, yielded to online platforms thanks to the various opportunities offered by the internet and social media. In addition to purchasing products via the internet and social media, consumers can also access all necessary information about a product effortlessly and can compare different products at the same time (Çelebi & Pırnar, 2017: 367).

Considering the studies relating to social media and pretension available in the literature, Sabuncuoğlu in his study (2015) evaluated the identity representations of individuals who use social media platforms as a new medium for the purpose of conspicuous consumption, which brands they prefer in order to show off, how and for what purpose they make these efforts and the reflections of their shares on other individuals. Taylor & Strutton (2016) showed how the use of Facebook is positively correlated with envy and narcissism. Eventually, they determined that an individuals'desire for self-promotion and tendency to participate in conspicuous consumption increased through the use of social media. In their study, Bayuk and Öz (2018) found that those who use social media more intensively are more prone to share pretentiously on such platforms. Besides, they concluded that the most fundamental and powerful motives of users who perform conspicuous consumption on social media is first imitation, then display and uniqueness, respectively. Taylor's study (2016) suggested that higher levels of social media use increases users' concerns about their lack of opportunity for socializing and gaining new experiences. It has been proven that negative emotions such as jealousy have a strong impact on motivating social media users to participate in the consumption of attention-grabbing status products and services. In particular FOMO (fear of missing out) was found to be motivating social media users to engage with prominent goods and experiences. The study revealed FOMO's relationship to those who share their photos with prestigious brands and products on social media and participate in luxurious vacations or status-raising activities.

### *2.3. Purchasing Intention (PI)*

Attitudes, intentions and behaviors are concepts that affect each other and generally occur in a certain order. According to the Theory of Planned Behavior, one's behavior is not realized only by one's own will; beliefs, attitudes, subjective norms and perceived behavioral control also determine the intention of the person to perform the behavior, and thus form the behavior. According to the theory, behavioral intention is the most important determinant and direct precursor of behavior (Ajzen, 1991). A consumer's plans to purchase a certain amount of a particular brand or product over a specified period of time is referred to as the consumer's intention to purchase a particular product or service. Purchasing intention does not necessarily mean that it will become a purchasing decision. Therefore, although it is difficult to measure, purchasing intention can be assessed within the framework of a consumer's willingness to purchase a product in the future (Çetin & Kumkale, 2016: 92). For marketing, it is very important to understand the value creation processes of enterprises that drive consumers to create purchasing intentions. It is because this understanding undoubtedly enables enterprises to see the transformation of their marketing activities such as profitability and market share (Zeren & Gokdagli, 2017: 92).

The studies (Park et al. 2007) on similar subjects indicate that purchasing intention increases when the quality and number of product reviews on online platforms by highly involved consumers increase. The studies (Jiang et al, 2010) revealed that consumers' greater interest and participation in the website positively affects purchasing intention and the studies (Hollebeek et al, 2007) show that consumers with high levels of product participation attach less importance to price, while consumers with less purchase participation give more importance to price reductions.

Among the studies in the literature on purchasing intention, the study by Çiftyıldız and Sütütemiz (2011) on car brands showed that the impact of some traits (hedonism for Mercedes, uniqueness for BMW and appeal for AUDI) vary according to the brand. The study concluded that the impact of brand features such as appeal and quality on eagerness to purchase can vary from brand to brand. In their study, Cop and Gülez (2016) revealed that the attitudes of female consumer generations towards imitation products, as well as seeking normative and informative sensitivity and innovation, personal taste, perception of price-quality, brand sensitivity and brand loyalty positively influenced the purchasing intention. Çetin and Kumkale (2016) demonstrated that when social media users see the selective and distinctive elements regarding a product they want to acquire on social media platforms, they develop a purchasing intention even if they do not spend a lot of time on social media. Can (2017) found that attitudes towards expressing appreciation and perceived herd behavior had a positive effect on purchasing intention. Orel et al. (2017) found that sentimental value is the strongest consumption value affecting a consumers' intention to purchase functional foods. This value was found to be followed respectively by novelty value, functional value-quality and situational value. They determined that purchasing intention for functional foods has a positive impact on purchasing behavior.

Aksoy & Gür (2018) studied the impact of consumer perception regarding social media commercials on purchasing intention. According to the results of the study, among a set of consumer perceptions regarding social media commercials, offering entertainment, trustworthiness and economic contributions were all found to have a positive impact on consumer's purchasing intention. In their study, Ural and Hallumoğlu (2018) revealed that, among consumption values, social and financial values have a positive and meaningful impact on the tendency to conspicuous consumption. However, they found that sentimental value and functional value have an insignificant impact on conspicuous consumption tendencies. In addition, they showed that among consumption values, social and financial values as well as conspicuous consumption tendencies have a positive impact on purchasing intention. However, they found that functional and sentimental values are irrelevant to purchasing intention. In their study, Ünal et al. (2019) showed that status consumption and creative selection affect social consumption motives, and clarity of self-conception prevents unpopular choices and similarities as well as affecting social consumption motives and purchasing intention. They suggested that peer pressure is a moderating factor in the relationship between clarity of self-conception and social consumption motives. They showed that while social consumption motives affect the attitude towards a luxury brand, positive attitudes increase the intention to purchase the luxury brand.

## 3. Hypotheses development

The research model developed during the literature review and tested during the study is shown in Figure 1. In this section, variables relating to the described research model and hypotheses based on the related literature have been developed.

As a result of the incredibly fast spread of laptops, tablets and especially smartphones following the introduction of the internet, almost everyone has begun to spend at least a few hours a day on social media platforms. A good number of people perform pretentious displays on these platforms. Indeed, it is a common sight to see the exhibition of, especially, luxury products and brands on social media to attract people (Bayuk & Öz, 2018: 2846). By enabling individuals to create a certain profile, social media offers people the opportunity to manifest and even boost their social status and personal image by sharing content symbolizing their own lifestyle (İlhan & Uğurhan, 2019: 38). One of the most important reasons for subjecting social media to conspicuous consumption is the desire to display status and prestige. In general, the lower classes in a society are motivated to imitate the higher social classes and to display luxury consumption due to the bandwagon effect. Therefore, they share their pretentious activities on social media platforms. The main purpose for such people is usually a feeling of belonging to and adapting to a group; because, if they succeed in doing so, they can achieve the status and prestige of being in a higher group. Again, higher class groups seek certain qualities such as uniqueness and rarity as an indicator of their privilege and exclusivity in their consumption/leisure activities performed due to the motives created by the snob effect. If they feel privileged, they share this on social media platforms to make their followers in the same or lower social groups jealous or to prove their power, status and prestige (Bayuk & Öz, 2018: 2849). On social media, people pretentiously share the privileged products they consume such as cars, houses, jewelry and clothing as well as privileged leisure activities such as fun vacations and private time spent at restaurants and cafes. There are a few principal reasons for this. First of all, since the identities on social media have a fluid structure, people do not consider themselves as one and only on these platforms. Therefore, the identities revealed on social media platforms do not always reflect reality. As a matter of fact, the individual displays himself on his social media account with a different, inflated, knowledgeable, sophisticated, wealthy, etc., personality due to his need for respect, appreciation or proof of belonging to a group etc.; thus, he can experience virtual satisfaction thanks to this mask. In other words, there may be significant differences between an individual's real image and the desired image (Sabuncuoğlu, 2015: 373).

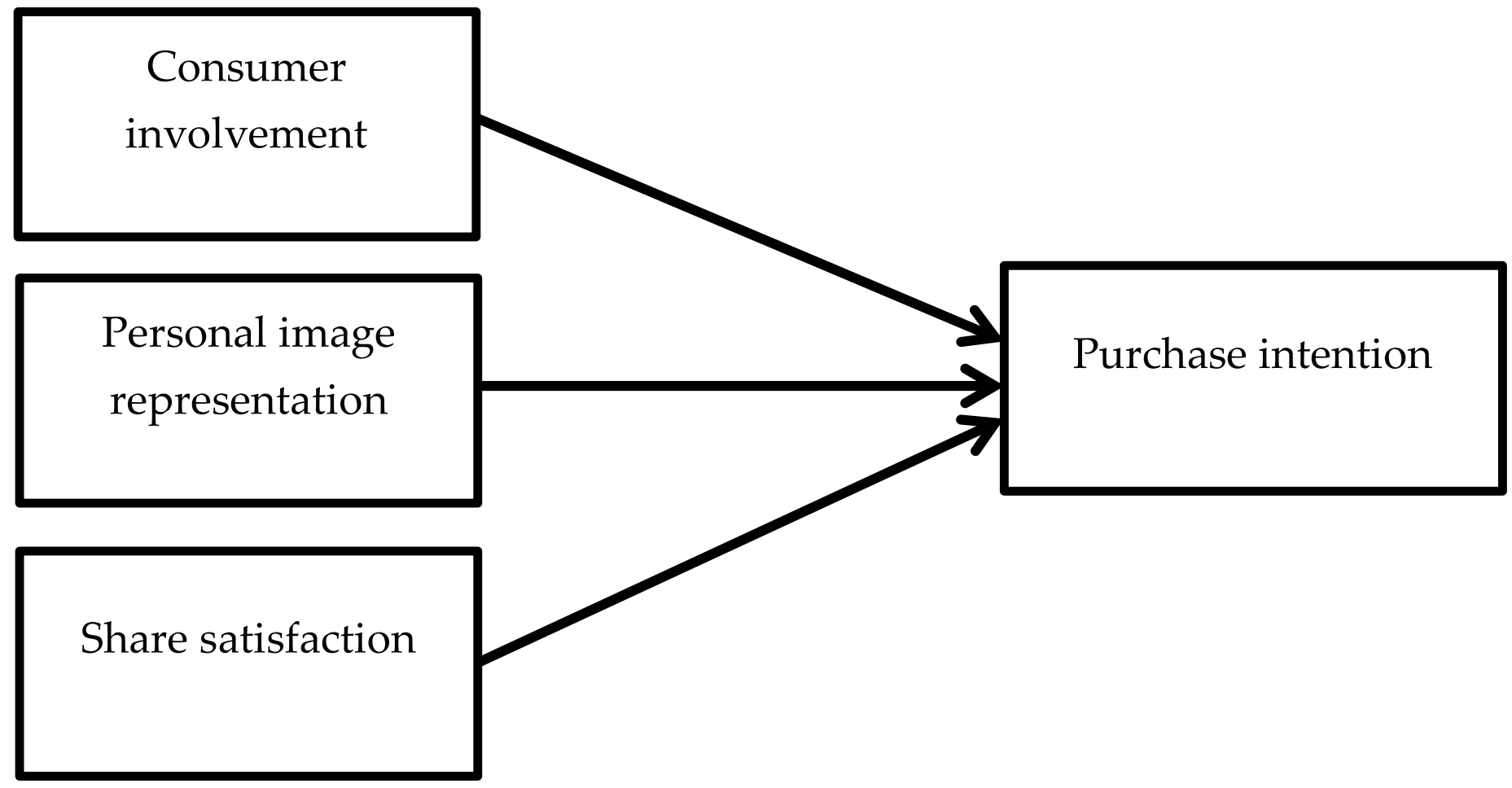

**Fig. 1.** Research Model

### *3.1. Consumer Involvement (CI)*

Involvement is the personal sense of importance and level of interest attributed to a stimulant under certain conditions. Thus, it is a motivational structure partially based on initial factors of personal values and needs (Zaichkowsky, 1986). Involvement is defined as the values that are important for the consumer as well as the personal level of interest in the situation or stimuli that leads to success. The expression of consumers' own experiences is explained as the extinction of behavior or the development of relations with the brand. Thus, Involvement means emphasis on the importance of the products owned, used and purchased by consumers. A consumer's interest is also expressed as the value perceived related to a certain situation or stimulus (Erbaş, 2016: 384). Involvement is an instinctual phenomenon that determines the level of motivation, stimulation and interest shown for a certain stimulus (Mitchell, 1981: 26). Consumer involvement is the emphasis on the importance of the products owned, used and purchased by the consumers. A consumer's desire for social popularity makes involvement easier through "likes" on social networks (Armağan & Gider, 2017: 696). Thus, the hypothesis of the study determined after a review of the literature to test the impact of consumer involvement on purchasing intention is as follows:

**$H_1$.** Consumer involvement positively affects purchasing intention.

### *3.2. Personal Image Representation (PIR)*

Social media offers individuals a platform where they can willingly manifest their lifestyles or access the lifestyles of others and can display their personal images. In addition, the fact that social media is free of time and space limitations, it is easy and cheap to access and even offers bidirectional communication making it suitable for conspicuous consumption (Sabuncuoğlu, 2015: 372). People try to create an image that they wish to exist in the minds of others because, an individual can exhibit his status in the society and can become different from the other people thanks to this image (Featherstone, 2013: 40–45). Consumers have begun to seek more than just physical satisfaction in the goods and services that they purchase for conspicuous consumption. Similarly, since consumers who have a tendency to conspicuous consumption want to experience such feelings in the goods and services they purchase, it could be suggested that sentimental value affects the tendency to conspicuous consumption (Ural & Hallumoğlu, 2018: 50). Therefore, the hypothesis of the study determined after a review of the literature to test the impact of personal image representation on purchasing intention is as follows:

**$H_2$.** Personal image representation positively affects purchasing intention.

*3.3. Share Satisfaction (SS)*

The feelings of a consumer created by a product play a role in each purchasing decision. The feelings created by the product such as enjoyment, satisfaction and relaxation etc. may affect conspicuous consumption tendency (Ural & Hallumoğlu, 2018: 50). Individuals make an effort to gain credit by sharing positive content or information about themselves on their profiles created on social media applications. If what is shared by the individual receives likes, the individual's self-confidence is accordingly boosted. This enables the individual to create a positive image in the eyes of other individuals. In this sense, there is an emphasis on the liking and commenting actions of individuals regarding conspicuous consumption tendencies on social media (Thoumrungroje, 2014: 8–11). In addition, it is suggested that the main purpose of individuals in social media applications is self-display and sharing their existence with others. Also, the literature discusses an individual's efforts to look like an important person and even a star in their social media shares (Uyanık, 2013: 2–4). Thus, the hypothesis of the study determined after a review of the literature to test the impact of share satisfaction on purchasing intention is as follows:

**$H_3$.** Share satisfaction positively affects purchasing intention.

## 4. Methodology

The study benefits from the scale developed by İlhan and Uğurhan (2019) to measure conspicuous consumption tendencies on social media as well as the study performed by Aksoy and Gür (2018) to measure purchasing intention. For this purpose, the first part of the survey consisted of 7 questions to measure demographic data. The second part covered 14 questions to measure conspicuous consumption on social media as well as 4 questions to measure purchasing intention. All statements in the second part have a 5-point Likert type design. These statements are in the format of a 1 to 5 points scale (1- Do not agree at all, 5- Completely agree). The study population is made up of social media users who live in the province of Gaziantep in Turkey. Due to time and cost constraints, the population was selected by the convenience sampling method which is one of the non-random sampling methods. When the population is 500,000 and greater and the confidence interval is 95%, a sufficient sample size is calculated as being 384 for the convenience sampling method (Yazıcıoğlu and Erdoğan, 2004: 50). Considering possible inconsistent and inadequate responses in the study, a face-to-face survey was carried out with 409 people. The surveys were evaluated on the basis of the data obtained from the study and the statistical tests were analyzed using the SPSS and AMOS package programs.

## 5. Results

Among 409 participants in the study, 44.3% (181) were male, and 55.7% (228) were female; thus, the gender distribution appears to have been relatively balanced. In addition, the majority of the participants were young; and 36.2% (148) of them were between the ages of 20 and 29. The highest level of education among the participants was high school. In addition, the monthly income distribution among the participants was similar. The majority of the study participants were engaged in their own business and the majority used social media for 1 to 2 hours per day. All the demographic data collected from the study is shown in Table 1.

**Table 1.** Demographic characteristics of the participants.

| | Frequency | Percentage |
|---|---|---|
| **Gender** | | |
| Male | 181 | 44.3 |
| Female | 228 | 55.7 |
| **Age** | | |
| ≤ 20 | 51 | 12.5 |
| 20-29 | 148 | 36.2 |
| 30-39 | 64 | 15.6 |
| 40-49 | 35 | 8.6 |
| 50-59 | 48 | 11.7 |
| ≥ 60 | 63 | 15.4 |

| **Marital Status** | | |
|---|---|---|
| Divorced/Widow | 47 | 11.5 |
| Single | 185 | 45.2 |
| Married | 177 | 43.3 |
| **Education** | | |
| Primary school | 71 | 17.4 |
| High school | 101 | 24.7 |
| Undergraduate | 68 | 16.6 |
| Graduate | 98 | 24.0 |
| Master | 50 | 12.2 |
| Doctorate | 21 | 5.1 |
| **Monthly Income** | | |
| ≤ 500 $ | 91 | 22.2 |
| 501 $-1000 $ | 68 | 16.6 |
| 1001 $-1500 $ | 90 | 22.0 |
| 1501 $-2000 $ | 63 | 15.4 |
| 2001 $-2500 $ | 70 | 17.1 |
| ≥2501 $ | 27 | 6.6 |
| **Job** | | |
| Unemployed | 24 | 5.9 |
| Student | 85 | 20.8 |
| Self-employed | 133 | 32.5 |
| Public servant/Academician | 82 | 20.0 |
| Private sector employee | 85 | 20.8 |
| **Daily social media usage** | | |
| Seldom | 95 | 23.2 |
| ≤ 1 h | 85 | 20.8 |
| 1 h – 2 h | 139 | 34.0 |
| 2 h – 3 h | 35 | 8.6 |
| ≥ 3 h | 55 | 13.4 |

### *5.1. Measurement model*

The reliability and validity of the scale were assessed in order to evaluate the measurement model. The two step approach suggested by Gerbing and Anderson (1988) for the measurement model construction and testing was followed. First, the measurement model to test the reliability was examined. AVE values were above 0.5 (Ruvio and Shogam, 2008; Fornell and Larcker, 1981), and CR were above 0.7 (Hair et al., 2006). Thus, average variance extracted (AVE), composite reliabilities (CR) and the Cronbach alpha coefficient were employed for the reliability evaluations of the scales in the survey. Cronbach's alpha of all constructs were found to be greater than the threshold of 0.7 (Kline, 2005) for basic research (Nunnally and Bernstein, 1994). AVE, CR and Cronbach's alpha reliability analyses were applied to the scale to determine whether the statements on the scale were consistent. The scales' AVE values vary between 0.556 and 0.713; the CR values vary between 0.799 and 0.869; and Cronbach's alpha reliability coefficients vary between 0.822 and 0.857. According to these findings, the scale in general and all variable groups can be considered as highly reliable. Table 2 presents the reliability analysis results of the main scale in the survey form. Secondly, a validity study was performed. The study was adapted from the models whose validity had been previously tested. Construct validity factor analysis of the applied survey form was determined by using the Direct Oblimin rotation method. According to the results of the factor analysis, Kaiser-Meyer-Olkin (KMO) value was determined as 0.955 and the Barlett test was found to be meaningful. Therefore, it was found that the scales used in the study were appropriate for factor analysis according to the results of the KMO and Barlett tests; and thus, it was possible to create meaningful data groups. In addition, all factor loadings were significant and above 0.6 (Bagozzi et al., 1991). In addition to all these, discriminant validity was examined by comparing the correlations among the constructs and the AVE values. As shown in Table 3, for each factor, the square root of AVE is significantly greater than its correlation coefficients with other factors, showing good discriminant validity (Gefen et al., 2000). Therefore, all the measures satisfied the discriminant validity of the constructs. The assessment of the validity results indicated that the constructs can be used to test the conceptual model.

**Table 2.** Standardized item loadings, AVE, CR and Alpha values.

| Variable | Item | Foctor loadings | AVE | CR | Cronbach's alpha |
|---|---|---|---|---|---|
| | | CFA | | | |
| Consumer Involvement | CI1 | 0.767 | 0.556 | 0.860 | 0.854 |
| | CI2 | 0.838 | | | |
| | CI3 | 0.827 | | | |
| | CI4 | 0.669 | | | |
| | CI5 | 0.600 | | | |
| Personal Image Representation | PIR1 | 0.750 | 0.573 | 0.869 | 0.857 |
| | PIR2 | 0.733 | | | |
| | PIR3 | 0.781 | | | |
| | PIR4 | 0.733 | | | |
| | PIR5 | 0.652 | | | |
| Share Satisfaction | SS1 | 0.808 | 0.713 | 0.799 | 0.822 |
| | SS2 | 0.788 | | | |
| | SS3 | 0.738 | | | |
| | SS4 | 0.467 | | | |
| Purchase Intention | PI1 | 0.826 | 0.624 | 0.868 | 0.852 |
| | PI2 | 0.788 | | | |
| | PI3 | 0.738 | | | |
| | PI4 | 0.667 | | | |

In addition, whether the data collected from the sample had a normal distribution was examined and the skewness and kurtosis values of the data were analyzed. If the values for skewness and kurtosis ranged between -2 to +2, it was assumed that the associated variable had a normal distribution (George and Mallery, 2010). The analyses showed that the data had a normal distribution.

**Table 3.** The square root of AVE (shown as bold at diagonal) factor correlation coefficients.

| | CI | PIR | SS | PI |
|---|---|---|---|---|
| Consumer Involvement | **0.746** | | | |
| Personal Image Representation | 0.543 | **0.757** | | |
| Share Satisfaction | -0.607 | - 0.551 | **0.844** | |
| Purchase Intention | -0.533 | -0.554 | 0.496 | **0.790** |

A structural equation analysis was conducted after satisfying the requirements of the measurement model. Table 4 lists the recommended and actual values of fit indices of the normalized fit index (NFI), the comparative fit index (CFI), and the root mean square error of approximation (RMSEA). For all indices, the actual values are better than the recommended values. Therefore, the model is a good fit.

**Table 4.** The recommended and actual values of fit indices.

| Fit index | chi2/df | CFI | NFI | RMSEA |
|---|---|---|---|---|
| Recommended values | <5.0* | ≥0.9** | ≥0.9*** | <0.08**** |
| Actual values | 3.982 | 0.919 | 0.935 | 0.052 |

Notes: *Bentler and Bonett (1980) **Hu and Bentler (1999) ***Fornell and Larcker(1981) ****Brown and Cudeck (1993)

### *5.2. Structural model*

While testing the study hypotheses, the structural equation model was utilized by using SPSS and AMOS. Whether the variables which were put forward first had sufficient model fit values was examined. These values were found to be within the appropriate range as shown in Table 4. The hypothesis tests examined the effect of "consumer involvement", "personal image representation" and "share satisfaction", which are the sub-dimensions of conspicuous consumption on social media, on purchasing intention. Table 5 lists the path coefficients and their significance. The results in Table 5 show that consumer involvement ($\beta=0.646$, $p\leq0.001$), personal image representation ($\beta=0.258$, $p\leq001$) and share satisfaction ($\beta=0.166$, $p\leq0.001$) were having a positive and significant impact on purchasing intention. Therefore, H1, H2 and H3 were supported.

**Table 5.** Hypothesis results.

| Structural paths | Hypothesis | Estimates | Supported or not |
|---|---|---|---|
| Consumer Involvement → Purchase Intention | H1 | 0.646*** | Yes |
| Personal Image Representation → Purchase Intention | H2 | 0.258*** | Yes |
| Share Satisfaction → Purchase Intention | H3 | 0.166*** | Yes |

Notes: ***p < 0.001

## 6. Discussion

In parallel with the spread of internet use in the last decade, social media has become indispensable in our lives. However, people have begun displaying themselves as being more prestigious and wealthy on social media; they even tend to leave an impression that they consume conspicuous products and services. In that sense, this study aims to determine whether a tendency to conspicuous consumption on social media shares affects purchasing intention. In this study, a research model was developed to examine the factors influencing purchasing intention; and the proposed model was tested with a structural equality model. Thus, the data received from the participants was analyzed and the results were interpreted.

### *6.1. Theoretical implications*

This study makes various theoretical contributions in terms of understanding the impact of consumption tendencies on social media on purchasing intention. First of all, consumer involvement positively affects purchasing intention in accordance with hypothesis $H_1$. This shows that social media sharing of products and services for which one has a high level of interest has an impact on purchasing intention. Positive motivation displayed towards a certain stimulus from social media shares gives purchasing ideas prominence. Accordingly, the hypothesis result resembles the findings in the studies of Armağan and Gider (2017), Çetin and Kumkale (2016), Koçak (2017), Orel et al. (2017) and Sherman et al. (1997). According to hypothesis $H_2$, personal image representation positively affects purchasing intention. Individuals began to display their lifestyle including their belongings, the places they visit, the foods they consume, and their leisure activities to other individuals through social media. Through such behavior on social media, individuals reflect their social status and the social class to which they belong. Therefore, individuals want to create an image which differentiates them from others. Accordingly, the hypothesis result resembles the findings in the studies of Bayuk and Öz (2018), İlhan and Uğurhan (2019), Ural and Hallumoğlu (2018) and Ünal et al. (2019). According to hypothesis $H_3$, share satisfaction positively affects purchasing intention. It was seen that when people's shares receive likes, people's self-confidence is accordingly boosted. This resembles the findings in the studies of Boğa and Başçı (2016), Can (2017), Koç (2015) and Sabuncuoğlu (2015), Taylor and Strutton (2016) and Van Dijck (2013).

### *6.2. Practical implications*

Veblen's *The Leisure Class Theory* has a similar content today; yet, it is brought to life through a new channel, which is social media. An individual's desire to define themselves as one who enjoys a luxury lifestyle and wealth as well as the realization of this effort as a representation through the symbolic meanings of brands have created new opportunities for luxury brands and companies. Based on the results of this experimental study, several managerial inferences have been obtained. The study has revealed that those who share on social media can consume to influence others or to show off, and are affected by each other. In particular luxury brands that want to increase the purchasing intentions of consumers can properly identify the personal interest levels of social media users and can develop their strategies accordingly. Consequently, companies interacting on social media with consumers can organize campaigns that stimulate social media shares by using popular venues; and thus, may positively affect the purchasing intention of individuals. In light of the findings of this study, companies are recommended to focus on professional shares on social media that are able to attract the attention of consumers and motivate them to share the information with their friends.

## 7. Conclusion

Consumption can be defined as satisfying human desires and needs with goods and services produced in exchange for a certain price. The meanings attributed to consumption have also changed over the years. Consumption has become an activity not only for meeting the biological needs of people, but also for people who want to have a place in society and acquire a certain status. Veblen argued that consumption has a conspicuous dimension and introduced the concept of conspicuous consumption to the literature. Consumers performing conspicuous consumption also benefit from the social status, prestige and reputation that the products/services have earned them by talking within their social circles about the products or services they use. Today social media is one of the practical platforms for conspicuous consumption. In this sense, other individuals following those involved in conspicuous consumption on social media are able to take them as examples and are influenced by the consumption behaviors of these leaders of opinion and are therefore able to imitate what they have. In addition, the study revealed that sharing in order to gain appreciation and social status is also observed in lower income groups.

### *7.1. Limitations and future research*

Since the sample used in the study consisted of social media users living in Gaziantep in Turkey, the results cannot be generalized to other social media users. Therefore, a study with a wider and more comprehensive sample should produce more detailed results and will be more useful. The following recommendations may be presented as suggestions to marketing academics working on social media, conspicuous consumption tendencies and consumer behavior as well as to many companies that interact with consumers on social media. In addition, a study could be carried out with participants using different social media platforms such as YouTube, Facebook, Instagram and Twitter. In this case, it would be possible to compare the trends in conspicuous consumption on social media among users of different social media platforms and also to analyze the purchasing intentions of consumers from multiple angles by employing a variety of disciplines.